\documentclass[aps,pre,reprint,superscriptaddress,amsmath]{revtex4-1}
\usepackage{lipsum}
\usepackage{graphicx}
\usepackage{subfigure}
\usepackage{textcomp}
\usepackage{setspace}
\begin{document}


\title{Comparison of two efficient methods for calculating partition functions}
\author{Le-Cheng Gong}
\affiliation{Institute of Modern Physics, Fudan University, Shanghai, 200433, China}
\affiliation{Applied Ion Beam Physics Laboratory, Fudan University, Shanghai, 200433, China}
\author{Bo-Yuan Ning}
\affiliation{Center for High Pressure Science $\&$ Technology Advanced Research, Shanghai, 202103, China}
\author{Tsu-Chien Weng}
\affiliation{Center for High Pressure Science $\&$ Technology Advanced Research, Shanghai, 202103, China}
\author{Xi-Jing Ning}
\email{Correspondence should be addressed to xjning@fudan.edu.cn}
\affiliation{Institute of Modern Physics, Fudan University, Shanghai, 200433, China}
\affiliation{Applied Ion Beam Physics Laboratory, Fudan University, Shanghai, 200433, China}

\date{\today}

\begin{abstract}
In the long-time pursuit of the solution to calculate the partition function (or free energy) of condensed matter,
Monte-Carlo-based nested sampling should be the state-of-the-art method, and very recently,
we established a direct integral approach that works at least four orders faster.
In present work, the above two methods were applied to solid argon at temperatures up to $300$K,
and the derived internal energy and pressure were compared with the molecular dynamics simulation
as well as experimental measurements,
showing that the calculation precision of our approach is about 10 times higher than that of the nested sampling method.
\end{abstract}

\maketitle
\section{Introduction}
\label{sec.intro}
The born of statistical physics laid a solid foundation to predict thermodynamic properties of macroscopic condensed matters.
Phase transitions~\cite{deterphase,hansen1969phase},
protein folding~\cite{proteinfolding} and
the optimal conditions for novel material growth could be predicted theoretically
as long as the partition function (PF) or free energy can be evaluated~\cite{chipot2007free}.
Nevertheless,
solutions to the PF has been a lon standing problem~\cite{jmoleliquids} and
attempts were reluctantly turned to the help of molecular simulations~\cite{states-based}.
With precedent efforts made in calculating the relative free energy,
e.g., Gibbs ensemble Monte Carlo (MC)~\cite{gibbsesemble} sampling and thermodynamic integration~\cite{ti},
more attentions have been paid to density of states (DOSs) for computing the absolute PF~\cite{metadynamics,hansmann1997parallel,wl,nxj}.
The Bayesian-statistics-based nested sampling (NS) may be the state-of-the-art one~\cite{skilling2004aip,skilling2006nested},
which aims at uniformly sampling a series of fixed fractions partitioned by potential energies in the configurational space to calculate DOS and
has been applied in several systems~\cite{deterphase,effcient,epmmixture,epmsolid,diffusivens,nsnvt,nssp,rapid,bolhuis2018nested,baldock2017constant,prx}.

Very recently,
we put forward a direct integral approach (DIA) to calculate the PF of condensed matters~\cite{nbyarxiv} and
the high accuracy has been proved by molecular dynamics~(MD) simulations of condensed copper and argon~\cite{nbyarxiv},
graphene and $\gamma$-graphyne materials~\cite{lyparxiv}, and silicene~\cite{lyparxiv2}.
Based on our reinterpreting the original sense of integral,
it was shown that DIA works at least four-order faster than NS~\cite{nbyarxiv}.
On the other hand,
it has not yet been confirmed whether the DIA has improved the computational precision of precedent MC methods.
In this work,
we carried out detailed analysis of DIA and NS in terms of the computational precision,
and performed MD simulations to test the precision of internal energy and equations of state derived from the PF.
It should be pointed out that
the tests with MD simulations, instead of experimental data,
is the most rigourous
because same interatomic potentials can be used in calculations of the PF and MD simulations,
which have been proved to be capable of producing very accurate results for various systems~\cite{constantpre/tem,canonical,constantpressure}.
If the results derived from PF are only compared with experimental measurements,
just as in most previous works,
it would yet be doubted that the method for calculating the PF is accurate or not
even if the agreements are excellent
since it would be very likely that
a deficient algorithm combined with an inappropriate empirical potential accidentally gives rise to an outcome close to the experiment.

The paper is organized as follows.
In Sec.~\ref{sec.method}, NS and DIA were briefly formulated,
and in Sec.~\ref{sec.res}, we first discussed the relationship between efficiency and accuracy of NS, and then,
performed MD simulations of solid argon to test the computational precision of DIA and NS,
showing that DIA has a much higher precision than NS.
In addition,
we found that NS works badly for the highly-condensed systems while DIA has no such a problem.
A comparison with experimental data of solid argon along the melting line was presented as well,
which further validates that DIA is more accurate than NS.

\section{Methods}
\label{sec.method}
PF is defined as a summation over the probabilities of all the microstates,
and for a canonical ensemble consisting of N particles confined in volume $V$ at temperature $T$,
it reads
\begin{equation}
\label{eq1}
\mathcal{Z}(N,V,T)=\frac{1}{N!\Lambda^{3N}}\int dq^{3N}\exp[-\beta U(q^{3N})],
\end{equation}
where $\Lambda$ is the thermal wavelength,
$\beta=1/k_BT$ with $k_B$ the Boltzmann constant,
$q^{3N}=\{q_1,q_2...,q_{3N}\}$ the Cartesian coordinates of particles and
$U(q^{3N})$ the potential energy.
The $3N$-dimensional integral on the right hand of Eq.(\ref{eq1}) is solely related to the microscopic states in configurational space,
the so-called \emph{configurational integral} (CI),
\begin{equation}
\label{eq2}
\mathcal{Q}=\int dq^{3N}\exp[-\beta U(q^{3N})].
\end{equation}

\subsection{Nested sampling}
In Eq.(\ref{eq2}),
microstates in configurational space are expressed in terms of coordinates of particles.
From another point of view,
we may also label the microstates by their corresponding potential energy and
the integral can be rewritten in terms of the DOS of potential energy
\begin{equation}
\label{eq3}
\mathcal{Q}=\frac{\int\exp(-\beta U)\Omega(U)dU}{\int\Omega(U)dU}
\end{equation}
where $\Omega(U)$ is the DOS of potential energy.

The strategy of NS is to partition the configurational space into a series of energy-decrease subdivisions numbered by $m$.
For the $m$th subspace with upper energy limit $U_m$,
a fixed number of configurations ($L$) with each energy $\varepsilon_i<U_m$ are generated by MC method
and ordered in a sequence as $\varepsilon_1<\varepsilon_2<\ldots<\varepsilon_L$.
The lower energy boundary $U_{m+1}$,
which is the upper one for the $(m+1)$th subspace,
is set to be the energy of a fixed fraction $\alpha$ of current subspace,
as $U_{m+1}=\varepsilon_I$ with $I=\alpha L$.
By the NS algorithm,
Eq.(\ref{eq3}) can be simplified as~\cite{effcient}
\begin{equation}
\label{eq3a}
\begin{split}
\mathcal{Q}&\approx\sum\limits_m\frac{\int_{U_{m+1}}^{U_{m}}\Omega(U)dU}{\int\Omega(U)dU}\exp(-\beta \langle U\rangle_m)\\
           &=\sum\limits_m\omega_m\exp(-\beta \langle U\rangle_m),
\end{split}
\end{equation}
where $ \langle U\rangle_m$ is an averaged energy of the $m$th subspace and
$\omega_m$ stands for the percentile of the $m$th phase space volume.
It is obvious that $\omega_m=\alpha^m-\alpha^{m+1}$,
and after $n$th iteration when the convergence condition is reached,
CI is evaluated as
\begin{equation}
\label{eq4}
\mathcal{Q}=\sum\limits^{n}_{m=1}(\alpha^{m}-\alpha^{m+1})\exp[-\beta(U_{m}+U_{m+1})/2],
\end{equation}
where $\langle U\rangle_m$ is chosen to be the arithmetic average of the boundary energies of each sampled partition~\cite{rapid,nsnvt}.
According to $E=-\frac{\partial\ln\mathcal{Z}}{\partial\beta}$,
the internal energy of the N-particle system is calculated by
\begin{widetext}
\begin{equation}
\label{eq5a}
E=\frac{3}{2}Nk_BT+\frac{\sum\limits^{n}_{m=1}[(U_{m}+U_{m+1})/2](\alpha^{m}-\alpha^{m+1})\exp[-\beta(U_{m}+U_{m+1})/2]}{\sum\limits^{n}_{m=1}(\alpha^{m}-\alpha^{m+1})\exp[-\beta(U_{m}+U_{m+1})/2]}.
\end{equation}
\end{widetext}
For determining the pressure by $P=\frac{1}{\beta}\frac{\partial\ln\mathcal{Z}}{\partial V}$,
another CI for the system with a volume of $V+\Delta V$ should be calculated and $P$ is obtained by
\begin{equation}
P\approx\frac{1}{\beta}\frac{1}{\mathcal{Q}(V)}\frac{\mathcal{Q}(V+\Delta V)-\mathcal{Q}(V)}{\Delta V}.
\label{eq5b}
\end{equation}

\subsection{Direct Integral Approach}
Consider Eq.(\ref{eq2}) and let the set $Q^{3N}=\{Q_1,Q_2...Q_{3N}\}$ be the coordinates of particles in the state of the lowest potential energy $U_0$,
we may introduce a function as
\begin{equation}
\label{eq7}
U'(q'^{3N})=U(q^{3N})-U_0,
\end{equation}
where $q'_i=q_i-Q_i$.
By inserting Eq.(\ref{eq7}) into Eq.(\ref{eq2}),
we obtain
\begin{equation}
\label{eq8}
\mathcal{Q}=e^{-\beta U_0}\int dq'^{3N}\exp[-\beta U'(q'^{3N})].
\end{equation}
According to our very recent work~\cite{nbyarxiv},
the integral can be solved as

\begin{equation}
\label{eq9}
Q=e^{-\beta U_0}\prod^{3N}_{i=1}\mathcal{L}_i,
\end{equation}
where $\mathcal{L}_i$ represents the effective length on the $i$th degree of freedom and is defined as
\begin{equation}
\label{eq10}
\mathcal{L}_i=\int e^{-\beta U'(0...q'_i...0)}dq'_i.
\end{equation}

For homogeneous systems with certain geometric symmetry,
such as perfect one-component crystals,
all the particles are equivalent and $U'$ felt by one particle moving along $q'_x$ may be the same as the one along $q'_y$ (or $q'_z$).
In such a case,
Eq.(\ref{eq9}) turns into
\begin{equation}
\label{eq11}
Q=e^{-\beta U_0}\mathcal{L}^{3N},
\end{equation}
where $\mathcal{L}$ is determined by Eq.(\ref{eq10}).
Otherwise,
it is needed to calculate the effective length, $\mathcal{L}_x$, $\mathcal{L}_y$, $\mathcal{L}_z$ respectively, and Eq.(\ref{eq9}) turns into
\begin{equation}
\label{eq12}
Q=e^{-\beta U_0}(\mathcal{L}_x\mathcal{L}_y\mathcal{L}_z)^N,
\end{equation}
and, $E$ and $P$ are thus evaluated as
\begin{widetext}
\begin{eqnarray}
E&=&\frac{3}{2}Nk_BT+U_0+3N\frac{\sum\limits_{i=1}^{n}U_i\exp[-\beta U_i]}{\sum\limits_{i=1}^{n}\exp[-\beta U_i]}, \label{eq13a}
\\
P&\approx&-\frac{U_0(V+\Delta V)-U_0(V)}{\Delta V}+\frac{3N}{\beta}\frac{1}{\mathcal{L}(V)}\frac{\mathcal{L}(V+\Delta V)-\mathcal{L}(V)}{\Delta V}.
\label{eq13b}
\end{eqnarray}
\end{widetext}


\begin{figure}
\includegraphics[width=3.2in,height=2.2in]{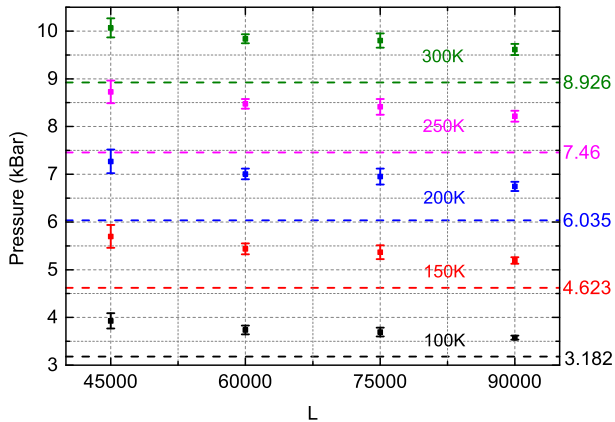}
\caption
{
\label{fig1}
(Color Online) The dependence of the pressure at different temperatures obtained by the NS and the standard deviations upon the $L$,
where the results of MD simulations are illustrated in dashed lines.
}
\end{figure}
\section{Comparisons and Discussions}
\label{sec.res}
The tested models are face-centered-cubic (FCC) solid Ar systems consisting of $500$ or $4000$ atoms
confined in a cubic box with different sizes,
and, NS and DIA were applied to calculate internal energy $E$ and pressure $P$ at different temperatures to be compared with MD simulations.
The interatomic potential for solid Ar was characterized by the commonly used pairwise 12-6 Lennard-Jones (L-J) potential~\cite{computersimu},
\begin{equation}
\label{eq14}
\phi (r_{ij})=4\epsilon[(\frac{\sigma}{r_{ij}})^{12}-(\frac{\sigma}{r_{ij}})^{6}],
\end{equation}
where $r_{ij}$ is the distance between atoms $i$ and $j$, $\epsilon=117.05$~(K), $\sigma=3.4$~{\AA} and the cutoff distance is $r_{cut}=12.0$~{\AA}.
The MD simulations with periodic boundary condition applied were performed by
the Large-scale Atomic/Molecular Massively Parallel Simulator software package~\cite{plimpton1995fast} with a time step of 0.1 fs.
The Nose-Hoover constant-temperature algorithm~\cite{nosehoover} was used to produce a canonical ensemble at temperature $T$.
The system was allowed to relax 20 ps at first and then continued to run for another 50 ps,
during which averages of $E$ and $P$ were recorded in every 10 fs.

To implement NS,
it should be at first to select appropriate values of $\alpha$ and $L$,
which cooperatively balance the computational efficiency and precision of $\mathcal {Q}$ [Eq.(\ref{eq4})].
Apparently, the larger the values of $\alpha$ and $L$ are,
the higher the calculation precision is, but the slower the computation speed is.
Although the initial choice of $\alpha$ made by P{\'a}rtay et al.~\cite{effcient} is $L/(L+1)$,
successive works~\cite{rapid,epmmixture,epmsolid,nssp} have showed that
a smaller value of $\alpha=1/2$ is sufficient enough to guarantee the calculation pricison and
enables NS to be applicable to systems consisting of up to several hundred atoms,
of which the computational cost is too expensive for NS with $\alpha=L/(L+1)$.
Therefore,
$\alpha=1/2$ was adopted in this work.
Cares should be also paid to the value of $L$ because,
besides the factors of efficiency and systematic errors mentioned above,
fluctuations of the calculated results in NS simulations closely depend on $L$ for a fixed $\alpha$~\cite{rapid}.
We performed NS with four different numbers of configuration ($L=45000, 60000, 75000, 90000$) to calculate the pressures of
the solid Ar system consisting of $500$ atoms with a density of $1.83$~g/cm$^3$ at different temperatures,
where the well-built cage model for solid systems~\cite{epmsolid} was used.
For each $L$,
we ran the NS simulations $15$ times to produce the averaged value of pressure which was compared with MD simulations to
see the relationship between the deviations and $L$.

As shown in Fig.\ref{fig1}~\cite{sm},
the pressures obtained by the NS are gradually approaching to those of the MD simulations as $L$ increases and
the corresponding fluctuations of NS is relatively larger with the smallest $L$.
On the other hand,
it should be noted that
the fluctuations does not monotonically decrease with the increase of $L$,
which was also observed in previous works~\cite{epmmixture}.
The fluctuations for $L=60000$ and $90000$ are almost the same,
which are about $30\%$ smaller than those with $L=75000$,
though the pressures with $L=90000$ are slightly closer to the MD simulations.
Considering that the computational time with $L=90000$ is twice as much as that with $L=60000$,
we chose $L=60000$ in the following work and
conducted the NS simulations at each $(N,V,T)$ conditions for 15 times to
calculate the averaged values of internal energy by Eq.(\ref{eq5a}) and pressure by Eq.(\ref{eq5b}),
where the volume difference $\Delta V$ was made by changing the length of the box by $1\%$ because our calculations showed that smaller volume difference would produce very unphysical results.
\begin{figure}
\includegraphics[width=3.5in,height=2.2in]{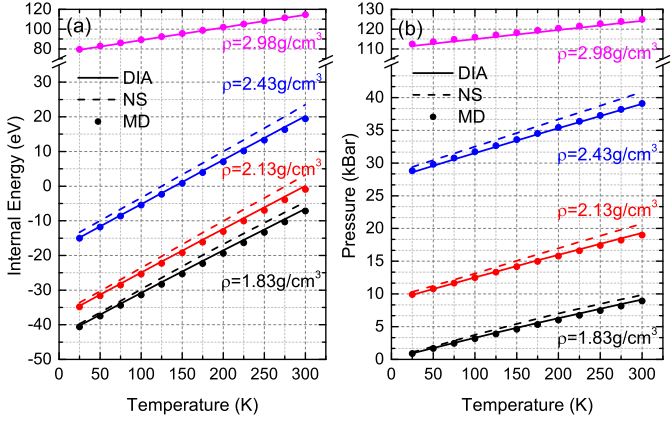}
\caption
{(Color Online) Internal energy (a) and pressure (b) of $500$ argon atoms in the solid state obtained by DIA (solid line), the NS (dashed line) and MD simulations (circles).
Different color stands for different density $\rho$.
}
\label{fig2}
\end{figure}

Relatively,
systematic parameters are much fewer for implementation of DIA.
For the solid Ar system,
the atoms were placed right at the FCC sites to produce $U_0$,
and $U'(0...q'_i...0)$ in Eq.(\ref{eq10}) was obtained by moving the center atoms along its $Z$-axis
($[100]$ direction) by 2{\AA}
while the coordinates of its $X$-axis, $Y$-axis, and of all the other particles were kept fixed.
$2\times10^4$ potential energies were recorded to calculate the $\mathcal{L}$ by Eq.(\ref{eq10}),
and, the internal energy and pressure were subsequently calculated by Eqs.(\ref{eq13a}) and (\ref{eq13b}),
where the volume difference was made by changing the length of the box by $10^{-3}\%$.

For the argon system of $500$ atoms with different densities ($1.83$, $2.13$, $2.43$ and $2.98$ g/cm$^3$)
at temperatures from $25$K to $300$K,
$E$ and $P$ obtained by DIA and the NS are shown in Fig.\ref{fig2},
where the corresponding quantities of $E_{MD}$ and $P_{MD}$ of MD simulations are also presented as comparisons~\cite{sm}.
For the systems with a density of $1.83$ g/cm$^3$ and $2.13$ g/cm$^3$,
the averaged relative difference of internal energy,
RDE ($=|\frac{E-E_{MD}}{E_{MD}}|$),
of DIA is less than $4.1\%$,
which is about four times smaller than that, $16.6\%$, of NS~\cite{NOTE}.
As the density increases up to $2.43$ g/cm$^3$ and $2.98$ g/cm$^3$,
the averaged $RDE$ of DIA decrease to $5.51\%$ and $0.48\%$ respectively,
while the averaged $RDE$ of the NS climbs up to $36.44\%$ for the density of $2.43$~g/cm$^3$ and
the NS fails to work for the system with density of $2.98$~g/cm$^3$.
As to precision of the pressure,
the averaged relative difference,
RDP ($=|\frac{P-P_{MD}}{P_{MD}}|$), of DIA is
$2.48\%$, $1.69\%$, $0.17\%$ and $0.63\%$ for the densities of $1.83$, $2.13$, $2.43$ and $2.98$~g/cm$^3$ respectively,
while the corresponding RDP of the NS is $10.22\%$, $9.18\%$, $4.54\%$ and $\infty$.

\begin{figure}
\includegraphics[width=3.5in,height=2.2in]{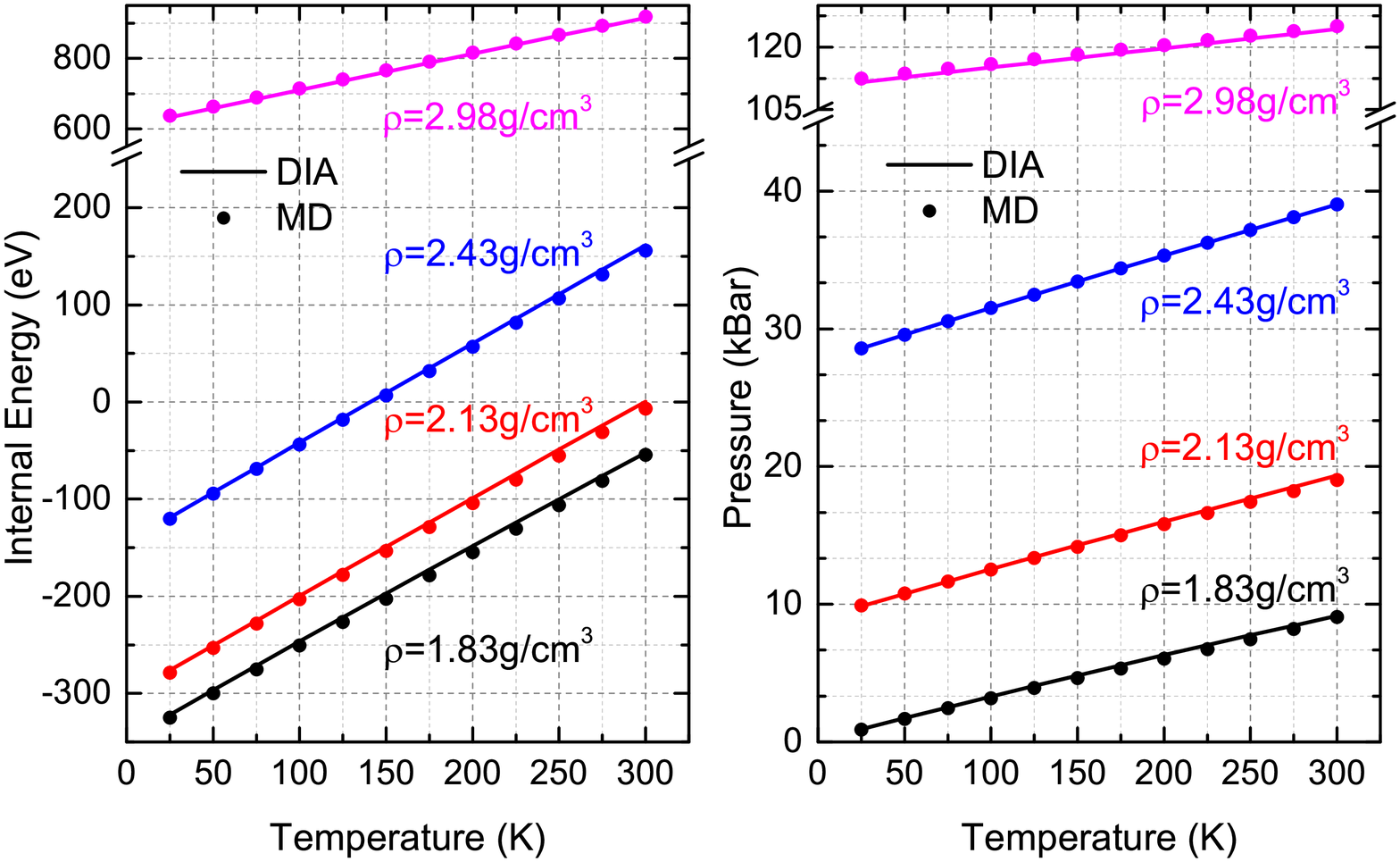}
\caption
{(Color Online) Internal energy (a) and pressure (b) of $4000$ argon atoms in the solid state obtained by DIA (solid line)
and MD simulations (circles).
Different color stands for different density $\rho$.}
\label{fig3}
\end{figure}

The above comparisons show that the calculation precision of DIA is much higher than that of the NS.
Furthermore,
DIA works better with increase of the density while the NS can hardly work when the density is higher than $2.98$~g/cm$^3$.
The difficulty should be attributed to numerical calculations of Eq.(\ref{eq4}),
where the factor ($\alpha^m-\alpha^{m+1}$) approaches to zero ($\alpha=1/2$) as $m$ approaches to larger number, meanwhile,
the factor $e^{-\beta(U_m+U_{m+1})/2}$ increases quickly when $U_m<0$,
which is the common case for the Ar systems with lower density and the product ($(\alpha^m-\alpha^{m+1})\cdot e^{-\beta (U_m+U_{m+1})/2}$) is not too large (or small) for the 16 bit number of computer to describe.
However, when the density is large enough that the $U_m>0$,
both ($\alpha^m-\alpha^{m+1}$) and ($e^{-\beta (U_m+U_{m+1})/2}$) approach to zero as $m$ getting larger,
and the product ($(\alpha^m-\alpha^{m+1})\cdot e^{-\beta (U_m+U_{m+1})/2}$) gets to be so small (but not exactly "0") that the output of computer is exact "0",
which makes the denominator in Eq.(\ref{eq5a}) be zero easily.
For this reason,
we failed to apply the NS to calculate $E$ and $P$ of the Ar system with a density of $2.98$~g/cm$^3$.
A larger value of $\alpha$ might be helpful while the computational efficiency would be slowed down.
By contrast,
DIA has no such a problem because the largest part of the potential energy, $U_0$ of the MSS,
has been extracted in Eqs.(\ref{eq7}) and (\ref{eq8}),
and the left part $U'$ is small enough to guarantee the precision of the integral for high density systems.

The lower precision for the NS calculating the pressure can be understood as follows.
The pressure is determined by Eq.(\ref{eq5b}), where the volume difference $\Delta V$ should be set as small as possible to achieve high precision.
However, the integral $\mathcal {Q}$ of Eq.(\ref{eq4}) is not very sensitive to the small changes of the volume $V$ because of the random characteristic of MC simulations,
leading to large fluctuations of $\mathcal {Q}(V+\Delta V)-\mathcal {Q}(V)$ for each running of the MC simulation. Our calculations showed that the large fluctuations would produce unphysical pressures when the $\Delta V$ is smaller than $10^{-4}\% V$, which corresponds to the length of the cubic box changed by $1\%$ adopted in our calculations.
In DIA for calculating the pressure [Eq.(\ref{eq13b})],
the involved quantities $U_0$ and $\mathcal{L}$ determined by Eq.(\ref{eq10}) are all sensitive to volume of the system, so the volume difference in Eq.(\ref{eq13b}) can be set much smaller.
We tried several values of the box length difference in the range of $10^{-1}\%-10^{-6}\%$ and
confirmed that the obtained pressures converges at the volume difference of $10^{-13}\%$ ($10^{-3}\%$ box length difference).

The computational efficiency of the NS and DIA depends on the number of the total potential calculation.
For the NS running, the MC algorithm has to work $6\times10^3-8\times10^3$ times each producing $60000$ configurations to reach the convergence,
so $3.6\times10^8-4.8\times10^8$ times of potential energy calculations must be performed to produce the $U_m$ in Eq.(\ref{eq4}).
Because of the fluctuations, the NS was run $15$ times for a given system to produce the averaged results,
thus the number of the total potential calculations is larger than $5.4\times10^9$, which is about five orders of magnitude larger than the one, $2\times10^4$, for running DIA in the same system.

Because of the ultra-high efficiency, DIA was applied to calculate the internal energy and pressure of solid argon composed of $4000$ atoms,
on which the NS costs too much computer hours and we have to give up the calculations, and we performed MD simulations to give comparisons.
As shown in Fig.\ref{fig3},
both $E$ and $P$ obtained by DIA coincide well with MD simulations
where both $RDE$ and $RDP$ of DIA are almost the same as those calculated in the $500$-atom system\cite{sm}.

\begin{figure}
\includegraphics[width=3.5in,height=2.2in]{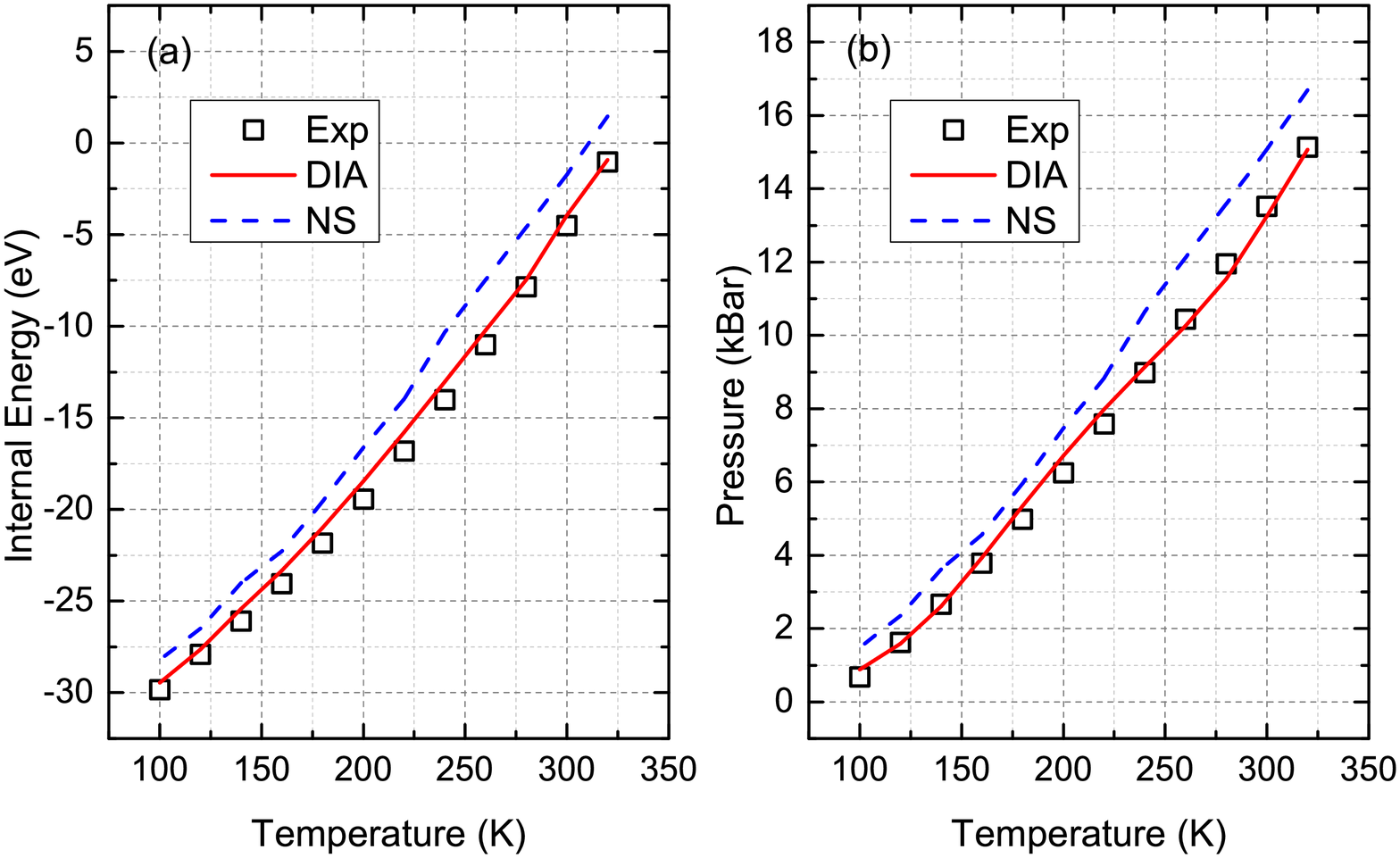}
\caption
{Internal energy (a) and pressure (b) of solid-state Ar,
from experiment (squares)~\cite{crawford1976thermodynamics}, DIA (solid lines), NS (dashed lines) along the melting line.
\label{fig4}
}
\end{figure}

Finally,
a comparison was made of DIA and the NS with experimental data of solid Ar along melting line~\cite{crawford1976thermodynamics}.
Considering the lower efficiency of the NS,
the simulated system for both DIA and NS consists of $500$ atoms and
the computational procedures are the same as described above.
As shown in Fig.\ref{fig4},
the internal energy and pressure obtained by DIA are significantly better than those of the NS.
The averaged relative deviation of internal energy and pressure to the experimental data is $5.34\%$ and $5.5\%$ for DIA,
which are about 6 times smaller than the ones, $39.12\%$ and $28.72\%$, for the NS.

\section{Conclusion}
\label{sec.con}
In summary,
by comparisons with MD simulations as well as experimental data,
we confirmed that the accuracy of DIA outperforms the NS.
The precision of DIA is about four times higher than that of NS for low-density systems
and about one order higher in high-density situations.
We also analyzed the intrinsic  deficiency of NS in calculations of systems under highly condensed situations.
Since the efficiency of DIA is at least five orders faster than that of the NS at the same time,
DIA paves a better way to investigate thermodynamic properties of condensed matters,
especially the ones with high density under extreme conditions.

\section{Acknowledgement}
TCW acknowledges the support by Nation Natural Science Foundation under Grant No.21727801.


\begin{thebibliography}{36}%
\makeatletter
\providecommand \@ifxundefined [1]{%
 \@ifx{#1\undefined}
}%
\providecommand \@ifnum [1]{%
 \ifnum #1\expandafter \@firstoftwo
 \else \expandafter \@secondoftwo
 \fi
}%
\providecommand \@ifx [1]{%
 \ifx #1\expandafter \@firstoftwo
 \else \expandafter \@secondoftwo
 \fi
}%
\providecommand \natexlab [1]{#1}%
\providecommand \enquote  [1]{``#1''}%
\providecommand \bibnamefont  [1]{#1}%
\providecommand \bibfnamefont [1]{#1}%
\providecommand \citenamefont [1]{#1}%
\providecommand \href@noop [0]{\@secondoftwo}%
\providecommand \href [0]{\begingroup \@sanitize@url \@href}%
\providecommand \@href[1]{\@@startlink{#1}\@@href}%
\providecommand \@@href[1]{\endgroup#1\@@endlink}%
\providecommand \@sanitize@url [0]{\catcode `\\12\catcode `\$12\catcode
  `\&12\catcode `\#12\catcode `\^12\catcode `\_12\catcode `\%12\relax}%
\providecommand \@@startlink[1]{}%
\providecommand \@@endlink[0]{}%
\providecommand \url  [0]{\begingroup\@sanitize@url \@url }%
\providecommand \@url [1]{\endgroup\@href {#1}{\urlprefix }}%
\providecommand \urlprefix  [0]{URL }%
\providecommand \Eprint [0]{\href }%
\providecommand \doibase [0]{http://dx.doi.org/}%
\providecommand \selectlanguage [0]{\@gobble}%
\providecommand \bibinfo  [0]{\@secondoftwo}%
\providecommand \bibfield  [0]{\@secondoftwo}%
\providecommand \translation [1]{[#1]}%
\providecommand \BibitemOpen [0]{}%
\providecommand \bibitemStop [0]{}%
\providecommand \bibitemNoStop [0]{.\EOS\space}%
\providecommand \EOS [0]{\spacefactor3000\relax}%
\providecommand \BibitemShut  [1]{\csname bibitem#1\endcsname}%
\let\auto@bib@innerbib\@empty
\bibitem [{\citenamefont {Baldock}\ \emph {et~al.}(2016)\citenamefont
  {Baldock}, \citenamefont {P{\'a}rtay}, \citenamefont {Bart{\'o}k},
  \citenamefont {Payne},\ and\ \citenamefont {Cs{\'a}nyi}}]{deterphase}%
  \BibitemOpen
  \bibfield  {author} {\bibinfo {author} {\bibfnamefont {R.~J.}\ \bibnamefont
  {Baldock}}, \bibinfo {author} {\bibfnamefont {L.~B.}\ \bibnamefont
  {P{\'a}rtay}}, \bibinfo {author} {\bibfnamefont {A.~P.}\ \bibnamefont
  {Bart{\'o}k}}, \bibinfo {author} {\bibfnamefont {M.~C.}\ \bibnamefont
  {Payne}}, \ and\ \bibinfo {author} {\bibfnamefont {G.}~\bibnamefont
  {Cs{\'a}nyi}},\ }\href@noop {} {\bibfield  {journal} {\bibinfo  {journal}
  {Physical Review B}\ }\textbf {\bibinfo {volume} {93}},\ \bibinfo {pages}
  {174108} (\bibinfo {year} {2016})}\BibitemShut {NoStop}%
\bibitem [{\citenamefont {Hansen}\ and\ \citenamefont
  {Verlet}(1969)}]{hansen1969phase}%
  \BibitemOpen
  \bibfield  {author} {\bibinfo {author} {\bibfnamefont {J.-P.}\ \bibnamefont
  {Hansen}}\ and\ \bibinfo {author} {\bibfnamefont {L.}~\bibnamefont
  {Verlet}},\ }\href@noop {} {\bibfield  {journal} {\bibinfo  {journal}
  {Physical Review}\ }\textbf {\bibinfo {volume} {184}},\ \bibinfo {pages}
  {151} (\bibinfo {year} {1969})}\BibitemShut {NoStop}%
\bibitem [{\citenamefont {Burkoff}\ \emph {et~al.}(2012)\citenamefont
  {Burkoff}, \citenamefont {V{\'a}rnai}, \citenamefont {Wells},\ and\
  \citenamefont {Wild}}]{proteinfolding}%
  \BibitemOpen
  \bibfield  {author} {\bibinfo {author} {\bibfnamefont {N.~S.}\ \bibnamefont
  {Burkoff}}, \bibinfo {author} {\bibfnamefont {C.}~\bibnamefont {V{\'a}rnai}},
  \bibinfo {author} {\bibfnamefont {S.~A.}\ \bibnamefont {Wells}}, \ and\
  \bibinfo {author} {\bibfnamefont {D.~L.}\ \bibnamefont {Wild}},\ }\href@noop
  {} {\bibfield  {journal} {\bibinfo  {journal} {Biophysical Journal}\ }\textbf
  {\bibinfo {volume} {102}},\ \bibinfo {pages} {878} (\bibinfo {year}
  {2012})}\BibitemShut {NoStop}%
\bibitem [{\citenamefont {Chipot}\ and\ \citenamefont
  {Pohorille}(2007)}]{chipot2007free}%
  \BibitemOpen
  \bibfield  {author} {\bibinfo {author} {\bibfnamefont {C.}~\bibnamefont
  {Chipot}}\ and\ \bibinfo {author} {\bibfnamefont {A.}~\bibnamefont
  {Pohorille}},\ }\href@noop {} {\emph {\bibinfo {title} {Free Energy
  Calculations: Theory and Applications in Chemistry and Biology}}},\
  Vol.~\bibinfo {volume} {86}\ (\bibinfo  {publisher} {Springer Science \&
  Business Media},\ \bibinfo {year} {2007})\BibitemShut {NoStop}%
\bibitem [{\citenamefont {Ushcats}\ \emph {et~al.}(2016)\citenamefont
  {Ushcats}, \citenamefont {Bulavin}, \citenamefont {Sysoev}, \citenamefont
  {Bardik},\ and\ \citenamefont {Alekseev}}]{jmoleliquids}%
  \BibitemOpen
  \bibfield  {author} {\bibinfo {author} {\bibfnamefont {M.~V.}\ \bibnamefont
  {Ushcats}}, \bibinfo {author} {\bibfnamefont {L.~A.}\ \bibnamefont
  {Bulavin}}, \bibinfo {author} {\bibfnamefont {V.~M.}\ \bibnamefont {Sysoev}},
  \bibinfo {author} {\bibfnamefont {V.~Y.}\ \bibnamefont {Bardik}}, \ and\
  \bibinfo {author} {\bibfnamefont {A.~N.}\ \bibnamefont {Alekseev}},\
  }\href@noop {} {\bibfield  {journal} {\bibinfo  {journal} {Journal of
  Molecular Liquids}\ }\textbf {\bibinfo {volume} {224}},\ \bibinfo {pages}
  {694} (\bibinfo {year} {2016})}\BibitemShut {NoStop}%
\bibitem [{\citenamefont {Singh}\ \emph {et~al.}(2012)\citenamefont {Singh},
  \citenamefont {Chopra},\ and\ \citenamefont {de~Pablo}}]{states-based}%
  \BibitemOpen
  \bibfield  {author} {\bibinfo {author} {\bibfnamefont {S.}~\bibnamefont
  {Singh}}, \bibinfo {author} {\bibfnamefont {M.}~\bibnamefont {Chopra}}, \
  and\ \bibinfo {author} {\bibfnamefont {J.~J.}\ \bibnamefont {de~Pablo}},\
  }\href@noop {} {\bibfield  {journal} {\bibinfo  {journal} {Annual Review of
  Chemical and Biomolecular Engineering}\ }\textbf {\bibinfo {volume} {3}},\
  \bibinfo {pages} {369} (\bibinfo {year} {2012})}\BibitemShut {NoStop}%
\bibitem [{\citenamefont {Mastny}\ and\ \citenamefont
  {de~Pablo}(2005)}]{gibbsesemble}%
  \BibitemOpen
  \bibfield  {author} {\bibinfo {author} {\bibfnamefont {E.~A.}\ \bibnamefont
  {Mastny}}\ and\ \bibinfo {author} {\bibfnamefont {J.~J.}\ \bibnamefont
  {de~Pablo}},\ }\href@noop {} {\bibfield  {journal} {\bibinfo  {journal} {The
  Journal of Chemical Physics}\ }\textbf {\bibinfo {volume} {122}},\ \bibinfo
  {pages} {124109} (\bibinfo {year} {2005})}\BibitemShut {NoStop}%
\bibitem [{\citenamefont {Mitchell}\ and\ \citenamefont {McCammon}(1991)}]{ti}%
  \BibitemOpen
  \bibfield  {author} {\bibinfo {author} {\bibfnamefont {M.~J.}\ \bibnamefont
  {Mitchell}}\ and\ \bibinfo {author} {\bibfnamefont {J.~A.}\ \bibnamefont
  {McCammon}},\ }\href@noop {} {\bibfield  {journal} {\bibinfo  {journal}
  {Journal of Computational Chemistry}\ }\textbf {\bibinfo {volume} {12}},\
  \bibinfo {pages} {271} (\bibinfo {year} {1991})}\BibitemShut {NoStop}%
\bibitem [{\citenamefont {Bussi}\ \emph {et~al.}(2006)\citenamefont {Bussi},
  \citenamefont {Laio},\ and\ \citenamefont {Parrinello}}]{metadynamics}%
  \BibitemOpen
  \bibfield  {author} {\bibinfo {author} {\bibfnamefont {G.}~\bibnamefont
  {Bussi}}, \bibinfo {author} {\bibfnamefont {A.}~\bibnamefont {Laio}}, \ and\
  \bibinfo {author} {\bibfnamefont {M.}~\bibnamefont {Parrinello}},\
  }\href@noop {} {\bibfield  {journal} {\bibinfo  {journal} {Physical Review
  Letters}\ }\textbf {\bibinfo {volume} {96}},\ \bibinfo {pages} {090601}
  (\bibinfo {year} {2006})}\BibitemShut {NoStop}%
\bibitem [{\citenamefont {Hansmann}(1997)}]{hansmann1997parallel}%
  \BibitemOpen
  \bibfield  {author} {\bibinfo {author} {\bibfnamefont {U.~H.}\ \bibnamefont
  {Hansmann}},\ }\href@noop {} {\bibfield  {journal} {\bibinfo  {journal}
  {Chemical Physics Letters}\ }\textbf {\bibinfo {volume} {281}},\ \bibinfo
  {pages} {140} (\bibinfo {year} {1997})}\BibitemShut {NoStop}%
\bibitem [{\citenamefont {Wang}\ and\ \citenamefont {Landau}(2001)}]{wl}%
  \BibitemOpen
  \bibfield  {author} {\bibinfo {author} {\bibfnamefont {F.}~\bibnamefont
  {Wang}}\ and\ \bibinfo {author} {\bibfnamefont {D.}~\bibnamefont {Landau}},\
  }\href@noop {} {\bibfield  {journal} {\bibinfo  {journal} {Physical Review
  Letters}\ }\textbf {\bibinfo {volume} {86}},\ \bibinfo {pages} {2050}
  (\bibinfo {year} {2001})}\BibitemShut {NoStop}%
\bibitem [{\citenamefont {Li}\ \emph {et~al.}(2016)\citenamefont {Li},
  \citenamefont {Ning}, \citenamefont {Zhuang},\ and\ \citenamefont
  {Ning}}]{nxj}%
  \BibitemOpen
  \bibfield  {author} {\bibinfo {author} {\bibfnamefont {J.-T.}\ \bibnamefont
  {Li}}, \bibinfo {author} {\bibfnamefont {B.-Y.}\ \bibnamefont {Ning}},
  \bibinfo {author} {\bibfnamefont {J.}~\bibnamefont {Zhuang}}, \ and\ \bibinfo
  {author} {\bibfnamefont {X.-J.}\ \bibnamefont {Ning}},\ }\href {\doibase
  10.1088/1674-1056/26/3/030501} {\bibfield  {journal} {\bibinfo  {journal}
  {Chinese Physics B}\ }\textbf {\bibinfo {volume} {26}},\ \bibinfo {pages}
  {030501} (\bibinfo {year} {2016})}\BibitemShut {NoStop}%
\bibitem [{\citenamefont {Skilling}(2004)}]{skilling2004aip}%
  \BibitemOpen
  \bibfield  {author} {\bibinfo {author} {\bibfnamefont {J.}~\bibnamefont
  {Skilling}},\ }\href@noop {} {\bibfield  {journal} {\bibinfo  {journal} {AIP
  Conference Proceedings}\ }\textbf {\bibinfo {volume} {735}},\ \bibinfo
  {pages} {395} (\bibinfo {year} {2004})}\BibitemShut {NoStop}%
\bibitem [{\citenamefont {Skilling}\ \emph {et~al.}(2006)\citenamefont
  {Skilling} \emph {et~al.}}]{skilling2006nested}%
  \BibitemOpen
  \bibfield  {author} {\bibinfo {author} {\bibfnamefont {J.}~\bibnamefont
  {Skilling}} \emph {et~al.},\ }\href@noop {} {\bibfield  {journal} {\bibinfo
  {journal} {Bayesian Analysis}\ }\textbf {\bibinfo {volume} {1}},\ \bibinfo
  {pages} {833} (\bibinfo {year} {2006})}\BibitemShut {NoStop}%
\bibitem [{\citenamefont {P{\'a}?rtay}\ \emph {et~al.}(2010)\citenamefont
  {P{\'a}?rtay}, \citenamefont {Bart{\'o}k},\ and\ \citenamefont
  {Cs{\'a}nyi}}]{effcient}%
  \BibitemOpen
  \bibfield  {author} {\bibinfo {author} {\bibfnamefont {L.~B.}\ \bibnamefont
  {P{\'a}?rtay}}, \bibinfo {author} {\bibfnamefont {A.~P.}\ \bibnamefont
  {Bart{\'o}k}}, \ and\ \bibinfo {author} {\bibfnamefont {G.}~\bibnamefont
  {Cs{\'a}nyi}},\ }\href@noop {} {\bibfield  {journal} {\bibinfo  {journal}
  {The Journal of Physical Chemistry B}\ }\textbf {\bibinfo {volume} {114}},\
  \bibinfo {pages} {10502} (\bibinfo {year} {2010})}\BibitemShut {NoStop}%
\bibitem [{\citenamefont {Do}\ \emph {et~al.}(2012)\citenamefont {Do},
  \citenamefont {Hirst},\ and\ \citenamefont {Wheatley}}]{epmmixture}%
  \BibitemOpen
  \bibfield  {author} {\bibinfo {author} {\bibfnamefont {H.}~\bibnamefont
  {Do}}, \bibinfo {author} {\bibfnamefont {J.~D.}\ \bibnamefont {Hirst}}, \
  and\ \bibinfo {author} {\bibfnamefont {R.~J.}\ \bibnamefont {Wheatley}},\
  }\href@noop {} {\bibfield  {journal} {\bibinfo  {journal} {The Journal of
  Physical Chemistry B}\ }\textbf {\bibinfo {volume} {116}},\ \bibinfo {pages}
  {4535} (\bibinfo {year} {2012})}\BibitemShut {NoStop}%
\bibitem [{\citenamefont {Do}\ and\ \citenamefont {Wheatley}(2012)}]{epmsolid}%
  \BibitemOpen
  \bibfield  {author} {\bibinfo {author} {\bibfnamefont {H.}~\bibnamefont
  {Do}}\ and\ \bibinfo {author} {\bibfnamefont {R.~J.}\ \bibnamefont
  {Wheatley}},\ }\href@noop {} {\bibfield  {journal} {\bibinfo  {journal}
  {Journal of Chemical Theory and Computation}\ }\textbf {\bibinfo {volume}
  {9}},\ \bibinfo {pages} {165} (\bibinfo {year} {2012})}\BibitemShut {NoStop}%
\bibitem [{\citenamefont {Brewer}\ \emph {et~al.}(2011)\citenamefont {Brewer},
  \citenamefont {P{\'a}rtay},\ and\ \citenamefont {Cs{\'a}nyi}}]{diffusivens}%
  \BibitemOpen
  \bibfield  {author} {\bibinfo {author} {\bibfnamefont {B.~J.}\ \bibnamefont
  {Brewer}}, \bibinfo {author} {\bibfnamefont {L.~B.}\ \bibnamefont
  {P{\'a}rtay}}, \ and\ \bibinfo {author} {\bibfnamefont {G.}~\bibnamefont
  {Cs{\'a}nyi}},\ }\href@noop {} {\bibfield  {journal} {\bibinfo  {journal}
  {Statistics and Computing}\ }\textbf {\bibinfo {volume} {21}},\ \bibinfo
  {pages} {649} (\bibinfo {year} {2011})}\BibitemShut {NoStop}%
\bibitem [{\citenamefont {Nielsen}(2013)}]{nsnvt}%
  \BibitemOpen
  \bibfield  {author} {\bibinfo {author} {\bibfnamefont {S.~O.}\ \bibnamefont
  {Nielsen}},\ }\href@noop {} {\bibfield  {journal} {\bibinfo  {journal} {The
  Journal of Chemical Physics}\ }\textbf {\bibinfo {volume} {139}},\ \bibinfo
  {pages} {124104} (\bibinfo {year} {2013})}\BibitemShut {NoStop}%
\bibitem [{\citenamefont {Wilson}\ \emph {et~al.}(2015)\citenamefont {Wilson},
  \citenamefont {Gelb},\ and\ \citenamefont {Nielsen}}]{nssp}%
  \BibitemOpen
  \bibfield  {author} {\bibinfo {author} {\bibfnamefont {B.~A.}\ \bibnamefont
  {Wilson}}, \bibinfo {author} {\bibfnamefont {L.~D.}\ \bibnamefont {Gelb}}, \
  and\ \bibinfo {author} {\bibfnamefont {S.~O.}\ \bibnamefont {Nielsen}},\
  }\href@noop {} {\bibfield  {journal} {\bibinfo  {journal} {The Journal of
  Chemical Physics}\ }\textbf {\bibinfo {volume} {143}},\ \bibinfo {pages}
  {154108} (\bibinfo {year} {2015})}\BibitemShut {NoStop}%
\bibitem [{\citenamefont {Do}\ \emph {et~al.}(2011)\citenamefont {Do},
  \citenamefont {Hirst},\ and\ \citenamefont {Wheatley}}]{rapid}%
  \BibitemOpen
  \bibfield  {author} {\bibinfo {author} {\bibfnamefont {H.}~\bibnamefont
  {Do}}, \bibinfo {author} {\bibfnamefont {J.~D.}\ \bibnamefont {Hirst}}, \
  and\ \bibinfo {author} {\bibfnamefont {R.~J.}\ \bibnamefont {Wheatley}},\
  }\href@noop {} {\bibfield  {journal} {\bibinfo  {journal} {The Journal of
  Chemical Physics}\ }\textbf {\bibinfo {volume} {135}},\ \bibinfo {pages}
  {174105} (\bibinfo {year} {2011})}\BibitemShut {NoStop}%
\bibitem [{\citenamefont {Bolhuis}\ and\ \citenamefont
  {Cs{\'a}nyi}(2018)}]{bolhuis2018nested}%
  \BibitemOpen
  \bibfield  {author} {\bibinfo {author} {\bibfnamefont {P.~G.}\ \bibnamefont
  {Bolhuis}}\ and\ \bibinfo {author} {\bibfnamefont {G.}~\bibnamefont
  {Cs{\'a}nyi}},\ }\href@noop {} {\bibfield  {journal} {\bibinfo  {journal}
  {Physical Review Letters}\ }\textbf {\bibinfo {volume} {120}},\ \bibinfo
  {pages} {250601} (\bibinfo {year} {2018})}\BibitemShut {NoStop}%
\bibitem [{\citenamefont {Baldock}\ \emph {et~al.}(2017)\citenamefont
  {Baldock}, \citenamefont {Bernstein}, \citenamefont {Salerno}, \citenamefont
  {P{\'a}rtay},\ and\ \citenamefont {Cs{\'a}nyi}}]{baldock2017constant}%
  \BibitemOpen
  \bibfield  {author} {\bibinfo {author} {\bibfnamefont {R.~J.}\ \bibnamefont
  {Baldock}}, \bibinfo {author} {\bibfnamefont {N.}~\bibnamefont {Bernstein}},
  \bibinfo {author} {\bibfnamefont {K.~M.}\ \bibnamefont {Salerno}}, \bibinfo
  {author} {\bibfnamefont {L.~B.}\ \bibnamefont {P{\'a}rtay}}, \ and\ \bibinfo
  {author} {\bibfnamefont {G.}~\bibnamefont {Cs{\'a}nyi}},\ }\href@noop {}
  {\bibfield  {journal} {\bibinfo  {journal} {Physical Review E}\ }\textbf
  {\bibinfo {volume} {96}},\ \bibinfo {pages} {043311} (\bibinfo {year}
  {2017})}\BibitemShut {NoStop}%
\bibitem [{\citenamefont {Martiniani}\ \emph {et~al.}(2014)\citenamefont
  {Martiniani}, \citenamefont {Stevenson}, \citenamefont {Wales},\ and\
  \citenamefont {Frenkel}}]{prx}%
  \BibitemOpen
  \bibfield  {author} {\bibinfo {author} {\bibfnamefont {S.}~\bibnamefont
  {Martiniani}}, \bibinfo {author} {\bibfnamefont {J.~D.}\ \bibnamefont
  {Stevenson}}, \bibinfo {author} {\bibfnamefont {D.~J.}\ \bibnamefont
  {Wales}}, \ and\ \bibinfo {author} {\bibfnamefont {D.}~\bibnamefont
  {Frenkel}},\ }\href@noop {} {\bibfield  {journal} {\bibinfo  {journal}
  {Physical Review X}\ }\textbf {\bibinfo {volume} {4}},\ \bibinfo {pages}
  {031034} (\bibinfo {year} {2014})}\BibitemShut {NoStop}%
\bibitem [{\citenamefont {Ning}\ \emph {et~al.}()\citenamefont {Ning},
  \citenamefont {Gong}, \citenamefont {Weng},\ and\ \citenamefont
  {Ning}}]{nbyarxiv}%
  \BibitemOpen
  \bibfield  {author} {\bibinfo {author} {\bibfnamefont {B.-Y.}\ \bibnamefont
  {Ning}}, \bibinfo {author} {\bibfnamefont {L.-C.}\ \bibnamefont {Gong}},
  \bibinfo {author} {\bibfnamefont {T.-C.}\ \bibnamefont {Weng}}, \ and\
  \bibinfo {author} {\bibfnamefont {X.-J.}\ \bibnamefont {Ning}},\ }\href@noop
  {} {\bibinfo  {journal} {arXiv:1901.08233}\ }\BibitemShut {NoStop}%
\bibitem [{\citenamefont {Liu}\ \emph {et~al.}({\natexlab{a}})\citenamefont
  {Liu}, \citenamefont {Ning}, \citenamefont {Gong}, \citenamefont {Weng},\
  and\ \citenamefont {Ning}}]{lyparxiv}%
  \BibitemOpen
\bibfield  {journal} {  }\bibfield  {author} {\bibinfo {author} {\bibfnamefont
  {Y.-P.}\ \bibnamefont {Liu}}, \bibinfo {author} {\bibfnamefont {B.-Y.}\
  \bibnamefont {Ning}}, \bibinfo {author} {\bibfnamefont {L.-C.}\ \bibnamefont
  {Gong}}, \bibinfo {author} {\bibfnamefont {T.-C.}\ \bibnamefont {Weng}}, \
  and\ \bibinfo {author} {\bibfnamefont {X.-J.}\ \bibnamefont {Ning}},\
  }\href@noop {} {\bibfield  {journal} {\bibinfo  {journal} {arXiv:1901.09205}\
  } ({\natexlab{a}})}\BibitemShut {NoStop}%
\bibitem [{\citenamefont {Liu}\ \emph {et~al.}({\natexlab{b}})\citenamefont
  {Liu}, \citenamefont {Ning}, \citenamefont {Gong}, \citenamefont {Weng},\
  and\ \citenamefont {Ning}}]{lyparxiv2}%
  \BibitemOpen
  \bibfield  {author} {\bibinfo {author} {\bibfnamefont {Y.-P.}\ \bibnamefont
  {Liu}}, \bibinfo {author} {\bibfnamefont {B.-Y.}\ \bibnamefont {Ning}},
  \bibinfo {author} {\bibfnamefont {L.-C.}\ \bibnamefont {Gong}}, \bibinfo
  {author} {\bibfnamefont {T.-C.}\ \bibnamefont {Weng}}, \ and\ \bibinfo
  {author} {\bibfnamefont {X.-J.}\ \bibnamefont {Ning}},\ }\href@noop {}
  {\bibfield  {journal} {\bibinfo  {journal} {arXiv:1902.06248}\ }
  ({\natexlab{b}})}\BibitemShut {NoStop}%
\bibitem [{\citenamefont {Andersen}(1980)}]{constantpre/tem}%
  \BibitemOpen
  \bibfield  {author} {\bibinfo {author} {\bibfnamefont {H.~C.}\ \bibnamefont
  {Andersen}},\ }\href@noop {} {\bibfield  {journal} {\bibinfo  {journal} {The
  Journal of Chemical Physics}\ }\textbf {\bibinfo {volume} {72}},\ \bibinfo
  {pages} {2384} (\bibinfo {year} {1980})}\BibitemShut {NoStop}%
\bibitem [{\citenamefont {Nos{\'e}}(1984)}]{canonical}%
  \BibitemOpen
  \bibfield  {author} {\bibinfo {author} {\bibfnamefont {S.}~\bibnamefont
  {Nos{\'e}}},\ }\href@noop {} {\bibfield  {journal} {\bibinfo  {journal}
  {Molecular Physics}\ }\textbf {\bibinfo {volume} {52}},\ \bibinfo {pages}
  {255} (\bibinfo {year} {1984})}\BibitemShut {NoStop}%
\bibitem [{\citenamefont {Hoover}(1986)}]{constantpressure}%
  \BibitemOpen
  \bibfield  {author} {\bibinfo {author} {\bibfnamefont {W.~G.}\ \bibnamefont
  {Hoover}},\ }\href@noop {} {\bibfield  {journal} {\bibinfo  {journal}
  {Physical Review A}\ }\textbf {\bibinfo {volume} {34}},\ \bibinfo {pages}
  {2499} (\bibinfo {year} {1986})}\BibitemShut {NoStop}%
\bibitem [{\citenamefont {Allen}\ and\ \citenamefont
  {Tildesley}(1989)}]{computersimu}%
  \BibitemOpen
  \bibfield  {author} {\bibinfo {author} {\bibfnamefont {M.~P.}\ \bibnamefont
  {Allen}}\ and\ \bibinfo {author} {\bibfnamefont {D.~J.}\ \bibnamefont
  {Tildesley}},\ }\href@noop {} {\emph {\bibinfo {title} {Computer simulation
  of liquids}}}\ (\bibinfo  {publisher} {Oxford university press},\ \bibinfo
  {year} {1989})\BibitemShut {NoStop}%
\bibitem [{\citenamefont {Plimpton}(1995)}]{plimpton1995fast}%
  \BibitemOpen
  \bibfield  {author} {\bibinfo {author} {\bibfnamefont {S.}~\bibnamefont
  {Plimpton}},\ }\href@noop {} {\bibfield  {journal} {\bibinfo  {journal}
  {Journal of Computational Physics}\ }\textbf {\bibinfo {volume} {117}},\
  \bibinfo {pages} {1} (\bibinfo {year} {1995})}\BibitemShut {NoStop}%
\bibitem [{\citenamefont {Evans}\ and\ \citenamefont
  {Holian}(1985)}]{nosehoover}%
  \BibitemOpen
  \bibfield  {author} {\bibinfo {author} {\bibfnamefont {D.~J.}\ \bibnamefont
  {Evans}}\ and\ \bibinfo {author} {\bibfnamefont {B.~L.}\ \bibnamefont
  {Holian}},\ }\href@noop {} {\bibfield  {journal} {\bibinfo  {journal} {The
  Journal of Chemical Physics}\ }\textbf {\bibinfo {volume} {83}},\ \bibinfo
  {pages} {4069} (\bibinfo {year} {1985})}\BibitemShut {NoStop}%
\bibitem [{sm()}]{sm}%
  \BibitemOpen
  \href@noop {} {\ }\bibinfo {note} {Detailed supporting data are shown in
  Supplementary Materials.}\BibitemShut {Stop}%
\bibitem [{NOT()}]{NOTE}%
  \BibitemOpen
  \href@noop {} {\ }\bibinfo {note} {For several given conditions, the relative
  differences of both DIA and NS between MD simulations are quite large, which
  might be due to the large fluctuations of MD simulations. For instance, at
  ($N=500$,$\rho=2.13$ g/cm$^3$,$T=300$K), the RDEs of DIA and NS are
  $107.59\%$ and $455.18\%$ respectively while the fluctuations of MD
  simulations of internal energy at this condition is $23.2\%$ with the
  $E_{MD}={-0.92}$~eV ($E_{DIA}=0.07$~eV and $E_{NS}=3.27$~eV). For a
  reasonable analysis, as a result, we excluded the data of which the MD
  fluctuations are over 5$\%$.}\BibitemShut {Stop}%
\bibitem [{\citenamefont {Crawford}\ \emph {et~al.}(1976)\citenamefont
  {Crawford}, \citenamefont {Lewis},\ and\ \citenamefont
  {Daniels}}]{crawford1976thermodynamics}%
  \BibitemOpen
  \bibfield  {author} {\bibinfo {author} {\bibfnamefont {R.}~\bibnamefont
  {Crawford}}, \bibinfo {author} {\bibfnamefont {W.}~\bibnamefont {Lewis}}, \
  and\ \bibinfo {author} {\bibfnamefont {W.}~\bibnamefont {Daniels}},\
  }\href@noop {} {\bibfield  {journal} {\bibinfo  {journal} {Journal of Physics
  C: Solid State Physics}\ }\textbf {\bibinfo {volume} {9}},\ \bibinfo {pages}
  {1381} (\bibinfo {year} {1976})}\BibitemShut {NoStop}%
\end{thebibliography}
%

\end{document}